\documentstyle[12pt]{article}
\begin{document}

\immediate\write16{<<WARNING: LINEDRAW macros work with emTeX-dvivers
		    and other drivers supporting emTeX \special's
		    (dviscr, dvihplj, dvidot, dvips, dviwin, etc.) >>}

\newdimen\Lengthunit	   \Lengthunit	= 1.5cm
\newcount\Nhalfperiods	   \Nhalfperiods= 9
\newcount\magnitude	   \magnitude = 1000

\catcode`\*=11
\newdimen\L*   \newdimen\d*   \newdimen\d**
\newdimen\dm*  \newdimen\dd*  \newdimen\dt*
\newdimen\a*   \newdimen\b*   \newdimen\c*
\newdimen\a**  \newdimen\b**
\newdimen\xL*  \newdimen\yL*
\newdimen\rx*  \newdimen\ry*
\newdimen\tmp* \newdimen\linwid*

\newcount\k*   \newcount\l*   \newcount\m*
\newcount\k**  \newcount\l**  \newcount\m**
\newcount\n*   \newcount\dn*  \newcount\r*
\newcount\N*   \newcount\*one \newcount\*two  \*one=1 \*two=2
\newcount\*ths \*ths=1000
\newcount\angle*  \newcount\q*	\newcount\q**
\newcount\angle** \angle**=0
\newcount\sc*	  \sc*=0

\newtoks\cos*  \cos*={1}
\newtoks\sin*  \sin*={0}

\catcode`\[=13

\def\rotate(#1){\advance\angle**#1\angle*=\angle**
\q**=\angle*\ifnum\q**<0\q**=-\q**\fi
\ifnum\q**>360\q*=\angle*\divide\q*360\multiply\q*360\advance\angle*-\q*\fi
\ifnum\angle*<0\advance\angle*360\fi\q**=\angle*\divide\q**90\q**=\q**
\def\sgcos*{+}\def\sgsin*{+}\relax
\ifcase\q**\or
 \def\sgcos*{-}\def\sgsin*{+}\or
 \def\sgcos*{-}\def\sgsin*{-}\or
 \def\sgcos*{+}\def\sgsin*{-}\else\fi
\q*=\q**
\multiply\q*90\advance\angle*-\q*
\ifnum\angle*>45\sc*=1\angle*=-\angle*\advance\angle*90\else\sc*=0\fi
\def[##1,##2]{\ifnum\sc*=0\relax
\edef\cs*{\sgcos*.##1}\edef\sn*{\sgsin*.##2}\ifcase\q**\or
 \edef\cs*{\sgcos*.##2}\edef\sn*{\sgsin*.##1}\or
 \edef\cs*{\sgcos*.##1}\edef\sn*{\sgsin*.##2}\or
 \edef\cs*{\sgcos*.##2}\edef\sn*{\sgsin*.##1}\else\fi\else
\edef\cs*{\sgcos*.##2}\edef\sn*{\sgsin*.##1}\ifcase\q**\or
 \edef\cs*{\sgcos*.##1}\edef\sn*{\sgsin*.##2}\or
 \edef\cs*{\sgcos*.##2}\edef\sn*{\sgsin*.##1}\or
 \edef\cs*{\sgcos*.##1}\edef\sn*{\sgsin*.##2}\else\fi\fi
\cos*={\cs*}\sin*={\sn*}\global\edef\gcos*{\cs*}\global\edef\gsin*{\sn*}}\relax
\ifcase\angle*[9999,0]\or
[999,017]\or[999,034]\or[998,052]\or[997,069]\or[996,087]\or
[994,104]\or[992,121]\or[990,139]\or[987,156]\or[984,173]\or
[981,190]\or[978,207]\or[974,224]\or[970,241]\or[965,258]\or
[961,275]\or[956,292]\or[951,309]\or[945,325]\or[939,342]\or
[933,358]\or[927,374]\or[920,390]\or[913,406]\or[906,422]\or
[898,438]\or[891,453]\or[882,469]\or[874,484]\or[866,499]\or
[857,515]\or[848,529]\or[838,544]\or[829,559]\or[819,573]\or
[809,587]\or[798,601]\or[788,615]\or[777,629]\or[766,642]\or
[754,656]\or[743,669]\or[731,681]\or[719,694]\or[707,707]\or
\else[9999,0]\fi}

\catcode`\[=12

\def\GRAPH(hsize=#1)#2{\hbox to #1\Lengthunit{#2\hss}}

\def\Linewidth#1{\global\linwid*=#1\relax
\global\divide\linwid*10\global\multiply\linwid*\mag
\global\divide\linwid*100\special{em:linewidth \the\linwid*}}

\Linewidth{.4pt}
\def\sm*{\special{em:moveto}}
\def\sl*{\special{em:lineto}}
\let\moveto=\sm*
\let\lineto=\sl*
\newbox\spm*   \newbox\spl*
\setbox\spm*\hbox{\sm*}
\setbox\spl*\hbox{\sl*}

\def\mov#1(#2,#3)#4{\rlap{\L*=#1\Lengthunit
\xL*=#2\L* \yL*=#3\L*
\xL*=\xscale\xL* \yL*=\yscale\yL*
\rx* \the\cos*\xL* \tmp* \the\sin*\yL* \advance\rx*-\tmp*
\ry* \the\cos*\yL* \tmp* \the\sin*\xL* \advance\ry*\tmp*
\kern\rx*\raise\ry*\hbox{#4}}}

\def\rmov*(#1,#2)#3{\rlap{\xL*=#1\yL*=#2\relax
\rx* \the\cos*\xL* \tmp* \the\sin*\yL* \advance\rx*-\tmp*
\ry* \the\cos*\yL* \tmp* \the\sin*\xL* \advance\ry*\tmp*
\kern\rx*\raise\ry*\hbox{#3}}}

\def\lin#1(#2,#3){\rlap{\sm*\mov#1(#2,#3){\sl*}}}

\def\arr*(#1,#2,#3){\rmov*(#1\dd*,#1\dt*){\sm*
\rmov*(#2\dd*,#2\dt*){\rmov*(#3\dt*,-#3\dd*){\sl*}}\sm*
\rmov*(#2\dd*,#2\dt*){\rmov*(-#3\dt*,#3\dd*){\sl*}}}}

\def\arrow#1(#2,#3){\rlap{\lin#1(#2,#3)\mov#1(#2,#3){\relax
\d**=-.012\Lengthunit\dd*=#2\d**\dt*=#3\d**
\arr*(1,10,4)\arr*(3,8,4)\arr*(4.8,4.2,3)}}}

\def\arrlin#1(#2,#3){\rlap{\L*=#1\Lengthunit\L*=.5\L*
\lin#1(#2,#3)\rmov*(#2\L*,#3\L*){\arrow.1(#2,#3)}}}

\def\dasharrow#1(#2,#3){\rlap{{\Lengthunit=0.9\Lengthunit
\dashlin#1(#2,#3)\mov#1(#2,#3){\sm*}}\mov#1(#2,#3){\sl*
\d**=-.012\Lengthunit\dd*=#2\d**\dt*=#3\d**
\arr*(1,10,4)\arr*(3,8,4)\arr*(4.8,4.2,3)}}}

\def\clap#1{\hbox to 0pt{\hss #1\hss}}

\def\ind(#1,#2)#3{\rlap{\L*=.1\Lengthunit
\xL*=#1\L* \yL*=#2\L*
\rx* \the\cos*\xL* \tmp* \the\sin*\yL* \advance\rx*-\tmp*
\ry* \the\cos*\yL* \tmp* \the\sin*\xL* \advance\ry*\tmp*
\kern\rx*\raise\ry*\hbox{\lower2pt\clap{$#3$}}}}

\def\sh*(#1,#2)#3{\rlap{\dm*=\the\n*\d**
\xL*=\xscale\dm* \yL*=\yscale\dm* \xL*=#1\xL* \yL*=#2\yL*
\rx* \the\cos*\xL* \tmp* \the\sin*\yL* \advance\rx*-\tmp*
\ry* \the\cos*\yL* \tmp* \the\sin*\xL* \advance\ry*\tmp*
\kern\rx*\raise\ry*\hbox{#3}}}

\def\calcnum*#1(#2,#3){\a*=1000sp\b*=1000sp\a*=#2\a*\b*=#3\b*
\ifdim\a*<0pt\a*-\a*\fi\ifdim\b*<0pt\b*-\b*\fi
\ifdim\a*>\b*\c*=.96\a*\advance\c*.4\b*
\else\c*=.96\b*\advance\c*.4\a*\fi
\k*\a*\multiply\k*\k*\l*\b*\multiply\l*\l*
\m*\k*\advance\m*\l*\n*\c*\r*\n*\multiply\n*\n*
\dn*\m*\advance\dn*-\n*\divide\dn*2\divide\dn*\r*
\advance\r*\dn*
\c*=\the\Nhalfperiods5sp\c*=#1\c*\ifdim\c*<0pt\c*-\c*\fi
\multiply\c*\r*\N*\c*\divide\N*10000}

\def\dashlin#1(#2,#3){\rlap{\calcnum*#1(#2,#3)\relax
\d**=#1\Lengthunit\ifdim\d**<0pt\d**-\d**\fi
\divide\N*2\multiply\N*2\advance\N*\*one
\divide\d**\N*\sm*\n*\*one\sh*(#2,#3){\sl*}\loop
\advance\n*\*one\sh*(#2,#3){\sm*}\advance\n*\*one
\sh*(#2,#3){\sl*}\ifnum\n*<\N*\repeat}}

\def\dashdotlin#1(#2,#3){\rlap{\calcnum*#1(#2,#3)\relax
\d**=#1\Lengthunit\ifdim\d**<0pt\d**-\d**\fi
\divide\N*2\multiply\N*2\advance\N*1\multiply\N*2\relax
\divide\d**\N*\sm*\n*\*two\sh*(#2,#3){\sl*}\loop
\advance\n*\*one\sh*(#2,#3){\kern-1.48pt\lower.5pt\hbox{\rm.}}\relax
\advance\n*\*one\sh*(#2,#3){\sm*}\advance\n*\*two
\sh*(#2,#3){\sl*}\ifnum\n*<\N*\repeat}}

\def\shl*(#1,#2)#3{\kern#1#3\lower#2#3\hbox{\unhcopy\spl*}}

\def\trianglin#1(#2,#3){\rlap{\toks0={#2}\toks1={#3}\calcnum*#1(#2,#3)\relax
\dd*=.57\Lengthunit\dd*=#1\dd*\divide\dd*\N*
\divide\dd*\*ths \multiply\dd*\magnitude
\d**=#1\Lengthunit\ifdim\d**<0pt\d**-\d**\fi
\multiply\N*2\divide\d**\N*\sm*\n*\*one\loop
\shl**{\dd*}\dd*-\dd*\advance\n*2\relax
\ifnum\n*<\N*\repeat\n*\N*\shl**{0pt}}}

\def\wavelin#1(#2,#3){\rlap{\toks0={#2}\toks1={#3}\calcnum*#1(#2,#3)\relax
\dd*=.23\Lengthunit\dd*=#1\dd*\divide\dd*\N*
\divide\dd*\*ths \multiply\dd*\magnitude
\d**=#1\Lengthunit\ifdim\d**<0pt\d**-\d**\fi
\multiply\N*4\divide\d**\N*\sm*\n*\*one\loop
\shl**{\dd*}\dt*=1.3\dd*\advance\n*\*one
\shl**{\dt*}\advance\n*\*one
\shl**{\dd*}\advance\n*\*two
\dd*-\dd*\ifnum\n*<\N*\repeat\n*\N*\shl**{0pt}}}

\def\w*lin(#1,#2){\rlap{\toks0={#1}\toks1={#2}\d**=\Lengthunit\dd*=-.12\d**
\divide\dd*\*ths \multiply\dd*\magnitude
\N*8\divide\d**\N*\sm*\n*\*one\loop
\shl**{\dd*}\dt*=1.3\dd*\advance\n*\*one
\shl**{\dt*}\advance\n*\*one
\shl**{\dd*}\advance\n*\*one
\shl**{0pt}\dd*-\dd*\advance\n*1\ifnum\n*<\N*\repeat}}

\def\l*arc(#1,#2)[#3][#4]{\rlap{\toks0={#1}\toks1={#2}\d**=\Lengthunit
\dd*=#3.037\d**\dd*=#4\dd*\dt*=#3.049\d**\dt*=#4\dt*\ifdim\d**>10mm\relax
\d**=.25\d**\n*\*one\shl**{-\dd*}\n*\*two\shl**{-\dt*}\n*3\relax
\shl**{-\dd*}\n*4\relax\shl**{0pt}\else
\ifdim\d**>5mm\d**=.5\d**\n*\*one\shl**{-\dt*}\n*\*two
\shl**{0pt}\else\n*\*one\shl**{0pt}\fi\fi}}

\def\d*arc(#1,#2)[#3][#4]{\rlap{\toks0={#1}\toks1={#2}\d**=\Lengthunit
\dd*=#3.037\d**\dd*=#4\dd*\d**=.25\d**\sm*\n*\*one\shl**{-\dd*}\relax
\n*3\relax\sh*(#1,#2){\xL*=\xscale\dd*\yL*=\yscale\dd*
\kern#2\xL*\lower#1\yL*\hbox{\sm*}}\n*4\relax\shl**{0pt}}}

\def\shl**#1{\c*=\the\n*\d**\d*=#1\relax
\a*=\the\toks0\c*\b*=\the\toks1\d*\advance\a*-\b*
\b*=\the\toks1\c*\d*=\the\toks0\d*\advance\b*\d*
\a*=\xscale\a*\b*=\yscale\b*
\rx* \the\cos*\a* \tmp* \the\sin*\b* \advance\rx*-\tmp*
\ry* \the\cos*\b* \tmp* \the\sin*\a* \advance\ry*\tmp*
\raise\ry*\rlap{\kern\rx*\unhcopy\spl*}}

\def\wlin*#1(#2,#3)[#4]{\rlap{\toks0={#2}\toks1={#3}\relax
\c*=#1\l*\c*\c*=.01\Lengthunit\m*\c*\divide\l*\m*
\c*=\the\Nhalfperiods5sp\multiply\c*\l*\N*\c*\divide\N*\*ths
\divide\N*2\multiply\N*2\advance\N*\*one
\dd*=.002\Lengthunit\dd*=#4\dd*\multiply\dd*\l*\divide\dd*\N*
\divide\dd*\*ths \multiply\dd*\magnitude
\d**=#1\multiply\N*4\divide\d**\N*\sm*\n*\*one\loop
\shl**{\dd*}\dt*=1.3\dd*\advance\n*\*one
\shl**{\dt*}\advance\n*\*one
\shl**{\dd*}\advance\n*\*two
\dd*-\dd*\ifnum\n*<\N*\repeat\n*\N*\shl**{0pt}}}

\def\wavebox#1{\setbox0\hbox{#1}\relax
\a*=\wd0\advance\a*14pt\b*=\ht0\advance\b*\dp0\advance\b*14pt\relax
\hbox{\kern9pt\relax
\rmov*(0pt,\ht0){\rmov*(-7pt,7pt){\wlin*\a*(1,0)[+]\wlin*\b*(0,-1)[-]}}\relax
\rmov*(\wd0,-\dp0){\rmov*(7pt,-7pt){\wlin*\a*(-1,0)[+]\wlin*\b*(0,1)[-]}}\relax
\box0\kern9pt}}

\def\rectangle#1(#2,#3){\relax
\lin#1(#2,0)\lin#1(0,#3)\mov#1(0,#3){\lin#1(#2,0)}\mov#1(#2,0){\lin#1(0,#3)}}

\def\dashrectangle#1(#2,#3){\dashlin#1(#2,0)\dashlin#1(0,#3)\relax
\mov#1(0,#3){\dashlin#1(#2,0)}\mov#1(#2,0){\dashlin#1(0,#3)}}

\def\waverectangle#1(#2,#3){\L*=#1\Lengthunit\a*=#2\L*\b*=#3\L*
\ifdim\a*<0pt\a*-\a*\def\x*{-1}\else\def\x*{1}\fi
\ifdim\b*<0pt\b*-\b*\def\y*{-1}\else\def\y*{1}\fi
\wlin*\a*(\x*,0)[-]\wlin*\b*(0,\y*)[+]\relax
\mov#1(0,#3){\wlin*\a*(\x*,0)[+]}\mov#1(#2,0){\wlin*\b*(0,\y*)[-]}}

\def\calcparab*{\ifnum\n*>\m*\k*\N*\advance\k*-\n*\else\k*\n*\fi
\a*=\the\k* sp\a*=10\a*\b*\dm*\advance\b*-\a*\k*\b*
\a*=\the\*ths\b*\divide\a*\l*\multiply\a*\k*
\divide\a*\l*\k*\*ths\r*\a*\advance\k*-\r*\dt*=\the\k*\L*}

\def\arcto#1(#2,#3)[#4]{\rlap{\toks0={#2}\toks1={#3}\calcnum*#1(#2,#3)\relax
\dm*=135sp\dm*=#1\dm*\d**=#1\Lengthunit\ifdim\dm*<0pt\dm*-\dm*\fi
\multiply\dm*\r*\a*=.3\dm*\a*=#4\a*\ifdim\a*<0pt\a*-\a*\fi
\advance\dm*\a*\N*\dm*\divide\N*10000\relax
\divide\N*2\multiply\N*2\advance\N*\*one
\L*=-.25\d**\L*=#4\L*\divide\d**\N*\divide\L*\*ths
\m*\N*\divide\m*2\dm*=\the\m*5sp\l*\dm*\sm*\n*\*one\loop
\calcparab*\shl**{-\dt*}\advance\n*1\ifnum\n*<\N*\repeat}}

\def\arrarcto#1(#2,#3)[#4]{\L*=#1\Lengthunit\L*=.54\L*
\arcto#1(#2,#3)[#4]\rmov*(#2\L*,#3\L*){\d*=.457\L*\d*=#4\d*\d**-\d*
\rmov*(#3\d**,#2\d*){\arrow.02(#2,#3)}}}

\def\dasharcto#1(#2,#3)[#4]{\rlap{\toks0={#2}\toks1={#3}\relax
\calcnum*#1(#2,#3)\dm*=\the\N*5sp\a*=.3\dm*\a*=#4\a*\ifdim\a*<0pt\a*-\a*\fi
\advance\dm*\a*\N*\dm*
\divide\N*20\multiply\N*2\advance\N*1\d**=#1\Lengthunit
\L*=-.25\d**\L*=#4\L*\divide\d**\N*\divide\L*\*ths
\m*\N*\divide\m*2\dm*=\the\m*5sp\l*\dm*
\sm*\n*\*one\loop\calcparab*
\shl**{-\dt*}\advance\n*1\ifnum\n*>\N*\else\calcparab*
\sh*(#2,#3){\xL*=#3\dt* \yL*=#2\dt*
\rx* \the\cos*\xL* \tmp* \the\sin*\yL* \advance\rx*\tmp*
\ry* \the\cos*\yL* \tmp* \the\sin*\xL* \advance\ry*-\tmp*
\kern\rx*\lower\ry*\hbox{\sm*}}\fi
\advance\n*1\ifnum\n*<\N*\repeat}}

\def\*shl*#1{\c*=\the\n*\d**\advance\c*#1\a**\d*\dt*\advance\d*#1\b**
\a*=\the\toks0\c*\b*=\the\toks1\d*\advance\a*-\b*
\b*=\the\toks1\c*\d*=\the\toks0\d*\advance\b*\d*
\rx* \the\cos*\a* \tmp* \the\sin*\b* \advance\rx*-\tmp*
\ry* \the\cos*\b* \tmp* \the\sin*\a* \advance\ry*\tmp*
\raise\ry*\rlap{\kern\rx*\unhcopy\spl*}}

\def\calcnormal*#1{\b**=10000sp\a**\b**\k*\n*\advance\k*-\m*
\multiply\a**\k*\divide\a**\m*\a**=#1\a**\ifdim\a**<0pt\a**-\a**\fi
\ifdim\a**>\b**\d*=.96\a**\advance\d*.4\b**
\else\d*=.96\b**\advance\d*.4\a**\fi
\d*=.01\d*\r*\d*\divide\a**\r*\divide\b**\r*
\ifnum\k*<0\a**-\a**\fi\d*=#1\d*\ifdim\d*<0pt\b**-\b**\fi
\k*\a**\a**=\the\k*\dd*\k*\b**\b**=\the\k*\dd*}

\def\wavearcto#1(#2,#3)[#4]{\rlap{\toks0={#2}\toks1={#3}\relax
\calcnum*#1(#2,#3)\c*=\the\N*5sp\a*=.4\c*\a*=#4\a*\ifdim\a*<0pt\a*-\a*\fi
\advance\c*\a*\N*\c*\divide\N*20\multiply\N*2\advance\N*-1\multiply\N*4\relax
\d**=#1\Lengthunit\dd*=.012\d**
\divide\dd*\*ths \multiply\dd*\magnitude
\ifdim\d**<0pt\d**-\d**\fi\L*=.25\d**
\divide\d**\N*\divide\dd*\N*\L*=#4\L*\divide\L*\*ths
\m*\N*\divide\m*2\dm*=\the\m*0sp\l*\dm*
\sm*\n*\*one\loop\calcnormal*{#4}\calcparab*
\*shl*{1}\advance\n*\*one\calcparab*
\*shl*{1.3}\advance\n*\*one\calcparab*
\*shl*{1}\advance\n*2\dd*-\dd*\ifnum\n*<\N*\repeat\n*\N*\shl**{0pt}}}

\def\triangarcto#1(#2,#3)[#4]{\rlap{\toks0={#2}\toks1={#3}\relax
\calcnum*#1(#2,#3)\c*=\the\N*5sp\a*=.4\c*\a*=#4\a*\ifdim\a*<0pt\a*-\a*\fi
\advance\c*\a*\N*\c*\divide\N*20\multiply\N*2\advance\N*-1\multiply\N*2\relax
\d**=#1\Lengthunit\dd*=.012\d**
\divide\dd*\*ths \multiply\dd*\magnitude
\ifdim\d**<0pt\d**-\d**\fi\L*=.25\d**
\divide\d**\N*\divide\dd*\N*\L*=#4\L*\divide\L*\*ths
\m*\N*\divide\m*2\dm*=\the\m*0sp\l*\dm*
\sm*\n*\*one\loop\calcnormal*{#4}\calcparab*
\*shl*{1}\advance\n*2\dd*-\dd*\ifnum\n*<\N*\repeat\n*\N*\shl**{0pt}}}

\def\hr*#1{\L*=\xscale\Lengthunit\ifnum
\angle**=0\clap{\vrule width#1\L* height.1pt}\else
\L*=#1\L*\L*=.5\L*\rmov*(-\L*,0pt){\sm*}\rmov*(\L*,0pt){\sl*}\fi}

\def\shade#1[#2]{\rlap{\Lengthunit=#1\Lengthunit
\special{em:linewidth .001pt}\relax
\mov(0,#2.05){\hr*{.994}}\mov(0,#2.1){\hr*{.980}}\relax
\mov(0,#2.15){\hr*{.953}}\mov(0,#2.2){\hr*{.916}}\relax
\mov(0,#2.25){\hr*{.867}}\mov(0,#2.3){\hr*{.798}}\relax
\mov(0,#2.35){\hr*{.715}}\mov(0,#2.4){\hr*{.603}}\relax
\mov(0,#2.45){\hr*{.435}}\special{em:linewidth \the\linwid*}}}

\def\dshade#1[#2]{\rlap{\special{em:linewidth .001pt}\relax
\Lengthunit=#1\Lengthunit\if#2-\def\t*{+}\else\def\t*{-}\fi
\mov(0,\t*.025){\relax
\mov(0,#2.05){\hr*{.995}}\mov(0,#2.1){\hr*{.988}}\relax
\mov(0,#2.15){\hr*{.969}}\mov(0,#2.2){\hr*{.937}}\relax
\mov(0,#2.25){\hr*{.893}}\mov(0,#2.3){\hr*{.836}}\relax
\mov(0,#2.35){\hr*{.760}}\mov(0,#2.4){\hr*{.662}}\relax
\mov(0,#2.45){\hr*{.531}}\mov(0,#2.5){\hr*{.320}}\relax
\special{em:linewidth \the\linwid*}}}}

\def\vdot{\rlap{\kern-1.9pt\lower1.8pt\hbox{$\scriptstyle\bullet$}}}
\def\vtimes{\rlap{\kern-3pt\lower1.8pt\hbox{$\scriptstyle\times$}}}
\def\vDot{\rlap{\kern-2.3pt\lower2.7pt\hbox{$\bullet$}}}
\def\vTimes{\rlap{\kern-3.6pt\lower2.4pt\hbox{$\times$}}}

\def\arc(#1)[#2,#3]{{\k*=#2\l*=#3\m*=\l*
\advance\m*-6\ifnum\k*>\l*\relax\else
{\rotate(#2)\mov(#1,0){\sm*}}\loop
\ifnum\k*<\m*\advance\k*5{\rotate(\k*)\mov(#1,0){\sl*}}\repeat
{\rotate(#3)\mov(#1,0){\sl*}}\fi}}

\def\dasharc(#1)[#2,#3]{{\k**=#2\n*=#3\advance\n*-1\advance\n*-\k**
\L*=1000sp\L*#1\L* \multiply\L*\n* \multiply\L*\Nhalfperiods
\divide\L*57\N*\L* \divide\N*2000\ifnum\N*=0\N*1\fi
\r*\n*	\divide\r*\N* \ifnum\r*<2\r*2\fi
\m**\r* \divide\m**2 \l**\r* \advance\l**-\m** \N*\n* \divide\N*\r*
\k**\r* \multiply\k**\N* \dn*\n* \advance\dn*-\k** \divide\dn*2\advance\dn*\*one
\r*\l** \divide\r*2\advance\dn*\r* \advance\N*-2\k**#2\relax
\ifnum\l**<6{\rotate(#2)\mov(#1,0){\sm*}}\advance\k**\dn*
{\rotate(\k**)\mov(#1,0){\sl*}}\advance\k**\m**
{\rotate(\k**)\mov(#1,0){\sm*}}\loop
\advance\k**\l**{\rotate(\k**)\mov(#1,0){\sl*}}\advance\k**\m**
{\rotate(\k**)\mov(#1,0){\sm*}}\advance\N*-1\ifnum\N*>0\repeat
{\rotate(#3)\mov(#1,0){\sl*}}\else\advance\k**\dn*
\arc(#1)[#2,\k**]\loop\advance\k**\m** \r*\k**
\advance\k**\l** {\arc(#1)[\r*,\k**]}\relax
\advance\N*-1\ifnum\N*>0\repeat
\advance\k**\m**\arc(#1)[\k**,#3]\fi}}

\def\triangarc#1(#2)[#3,#4]{{\k**=#3\n*=#4\advance\n*-\k**
\L*=1000sp\L*#2\L* \multiply\L*\n* \multiply\L*\Nhalfperiods
\divide\L*57\N*\L* \divide\N*1000\ifnum\N*=0\N*1\fi
\d**=#2\Lengthunit \d*\d** \divide\d*57\multiply\d*\n*
\r*\n*	\divide\r*\N* \ifnum\r*<2\r*2\fi
\m**\r* \divide\m**2 \l**\r* \advance\l**-\m** \N*\n* \divide\N*\r*
\dt*\d* \divide\dt*\N* \dt*.5\dt* \dt*#1\dt*
\divide\dt*1000\multiply\dt*\magnitude
\k**\r* \multiply\k**\N* \dn*\n* \advance\dn*-\k** \divide\dn*2\relax
\r*\l** \divide\r*2\advance\dn*\r* \advance\N*-1\k**#3\relax
{\rotate(#3)\mov(#2,0){\sm*}}\advance\k**\dn*
{\rotate(\k**)\mov(#2,0){\sl*}}\advance\k**-\m**\advance\l**\m**\loop\dt*-\dt*
\d*\d** \advance\d*\dt*
\advance\k**\l**{\rotate(\k**)\rmov*(\d*,0pt){\sl*}}%
\advance\N*-1\ifnum\N*>0\repeat\advance\k**\m**
{\rotate(\k**)\mov(#2,0){\sl*}}{\rotate(#4)\mov(#2,0){\sl*}}}}

\def\wavearc#1(#2)[#3,#4]{{\k**=#3\n*=#4\advance\n*-\k**
\L*=4000sp\L*#2\L* \multiply\L*\n* \multiply\L*\Nhalfperiods
\divide\L*57\N*\L* \divide\N*1000\ifnum\N*=0\N*1\fi
\d**=#2\Lengthunit \d*\d** \divide\d*57\multiply\d*\n*
\r*\n*	\divide\r*\N* \ifnum\r*=0\r*1\fi
\m**\r* \divide\m**2 \l**\r* \advance\l**-\m** \N*\n* \divide\N*\r*
\dt*\d* \divide\dt*\N* \dt*.7\dt* \dt*#1\dt*
\divide\dt*1000\multiply\dt*\magnitude
\k**\r* \multiply\k**\N* \dn*\n* \advance\dn*-\k** \divide\dn*2\relax
\divide\N*4\advance\N*-1\k**#3\relax
{\rotate(#3)\mov(#2,0){\sm*}}\advance\k**\dn*
{\rotate(\k**)\mov(#2,0){\sl*}}\advance\k**-\m**\advance\l**\m**\loop\dt*-\dt*
\d*\d** \advance\d*\dt* \dd*\d** \advance\dd*1.3\dt*
\advance\k**\r*{\rotate(\k**)\rmov*(\d*,0pt){\sl*}}\relax
\advance\k**\r*{\rotate(\k**)\rmov*(\dd*,0pt){\sl*}}\relax
\advance\k**\r*{\rotate(\k**)\rmov*(\d*,0pt){\sl*}}\relax
\advance\k**\r*
\advance\N*-1\ifnum\N*>0\repeat\advance\k**\m**
{\rotate(\k**)\mov(#2,0){\sl*}}{\rotate(#4)\mov(#2,0){\sl*}}}}

\def\gmov*#1(#2,#3)#4{\rlap{\L*=#1\Lengthunit
\xL*=#2\L* \yL*=#3\L*
\rx* \gcos*\xL* \tmp* \gsin*\yL* \advance\rx*-\tmp*
\ry* \gcos*\yL* \tmp* \gsin*\xL* \advance\ry*\tmp*
\rx*=\xscale\rx* \ry*=\yscale\ry*
\xL* \the\cos*\rx* \tmp* \the\sin*\ry* \advance\xL*-\tmp*
\yL* \the\cos*\ry* \tmp* \the\sin*\rx* \advance\yL*\tmp*
\kern\xL*\raise\yL*\hbox{#4}}}

\def\rgmov*(#1,#2)#3{\rlap{\xL*#1\yL*#2\relax
\rx* \gcos*\xL* \tmp* \gsin*\yL* \advance\rx*-\tmp*
\ry* \gcos*\yL* \tmp* \gsin*\xL* \advance\ry*\tmp*
\rx*=\xscale\rx* \ry*=\yscale\ry*
\xL* \the\cos*\rx* \tmp* \the\sin*\ry* \advance\xL*-\tmp*
\yL* \the\cos*\ry* \tmp* \the\sin*\rx* \advance\yL*\tmp*
\kern\xL*\raise\yL*\hbox{#3}}}

\def\Earc(#1)[#2,#3][#4,#5]{{\k*=#2\l*=#3\m*=\l*
\advance\m*-6\ifnum\k*>\l*\relax\else\def\xscale{#4}\def\yscale{#5}\relax
{\angle**0\rotate(#2)}\gmov*(#1,0){\sm*}\loop
\ifnum\k*<\m*\advance\k*5\relax
{\angle**0\rotate(\k*)}\gmov*(#1,0){\sl*}\repeat
{\angle**0\rotate(#3)}\gmov*(#1,0){\sl*}\relax
\def\xscale{1}\def\yscale{1}\fi}}

\def\dashEarc(#1)[#2,#3][#4,#5]{{\k**=#2\n*=#3\advance\n*-1\advance\n*-\k**
\L*=1000sp\L*#1\L* \multiply\L*\n* \multiply\L*\Nhalfperiods
\divide\L*57\N*\L* \divide\N*2000\ifnum\N*=0\N*1\fi
\r*\n*	\divide\r*\N* \ifnum\r*<2\r*2\fi
\m**\r* \divide\m**2 \l**\r* \advance\l**-\m** \N*\n* \divide\N*\r*
\k**\r*\multiply\k**\N* \dn*\n* \advance\dn*-\k** \divide\dn*2\advance\dn*\*one
\r*\l** \divide\r*2\advance\dn*\r* \advance\N*-2\k**#2\relax
\ifnum\l**<6\def\xscale{#4}\def\yscale{#5}\relax
{\angle**0\rotate(#2)}\gmov*(#1,0){\sm*}\advance\k**\dn*
{\angle**0\rotate(\k**)}\gmov*(#1,0){\sl*}\advance\k**\m**
{\angle**0\rotate(\k**)}\gmov*(#1,0){\sm*}\loop
\advance\k**\l**{\angle**0\rotate(\k**)}\gmov*(#1,0){\sl*}\advance\k**\m**
{\angle**0\rotate(\k**)}\gmov*(#1,0){\sm*}\advance\N*-1\ifnum\N*>0\repeat
{\angle**0\rotate(#3)}\gmov*(#1,0){\sl*}\def\xscale{1}\def\yscale{1}\else
\advance\k**\dn* \Earc(#1)[#2,\k**][#4,#5]\loop\advance\k**\m** \r*\k**
\advance\k**\l** {\Earc(#1)[\r*,\k**][#4,#5]}\relax
\advance\N*-1\ifnum\N*>0\repeat
\advance\k**\m**\Earc(#1)[\k**,#3][#4,#5]\fi}}

\def\triangEarc#1(#2)[#3,#4][#5,#6]{{\k**=#3\n*=#4\advance\n*-\k**
\L*=1000sp\L*#2\L* \multiply\L*\n* \multiply\L*\Nhalfperiods
\divide\L*57\N*\L* \divide\N*1000\ifnum\N*=0\N*1\fi
\d**=#2\Lengthunit \d*\d** \divide\d*57\multiply\d*\n*
\r*\n*	\divide\r*\N* \ifnum\r*<2\r*2\fi
\m**\r* \divide\m**2 \l**\r* \advance\l**-\m** \N*\n* \divide\N*\r*
\dt*\d* \divide\dt*\N* \dt*.5\dt* \dt*#1\dt*
\divide\dt*1000\multiply\dt*\magnitude
\k**\r* \multiply\k**\N* \dn*\n* \advance\dn*-\k** \divide\dn*2\relax
\r*\l** \divide\r*2\advance\dn*\r* \advance\N*-1\k**#3\relax
\def\xscale{#5}\def\yscale{#6}\relax
{\angle**0\rotate(#3)}\gmov*(#2,0){\sm*}\advance\k**\dn*
{\angle**0\rotate(\k**)}\gmov*(#2,0){\sl*}\advance\k**-\m**
\advance\l**\m**\loop\dt*-\dt* \d*\d** \advance\d*\dt*
\advance\k**\l**{\angle**0\rotate(\k**)}\rgmov*(\d*,0pt){\sl*}\relax
\advance\N*-1\ifnum\N*>0\repeat\advance\k**\m**
{\angle**0\rotate(\k**)}\gmov*(#2,0){\sl*}\relax
{\angle**0\rotate(#4)}\gmov*(#2,0){\sl*}\def\xscale{1}\def\yscale{1}}}

\def\waveEarc#1(#2)[#3,#4][#5,#6]{{\k**=#3\n*=#4\advance\n*-\k**
\L*=4000sp\L*#2\L* \multiply\L*\n* \multiply\L*\Nhalfperiods
\divide\L*57\N*\L* \divide\N*1000\ifnum\N*=0\N*1\fi
\d**=#2\Lengthunit \d*\d** \divide\d*57\multiply\d*\n*
\r*\n*	\divide\r*\N* \ifnum\r*=0\r*1\fi
\m**\r* \divide\m**2 \l**\r* \advance\l**-\m** \N*\n* \divide\N*\r*
\dt*\d* \divide\dt*\N* \dt*.7\dt* \dt*#1\dt*
\divide\dt*1000\multiply\dt*\magnitude
\k**\r* \multiply\k**\N* \dn*\n* \advance\dn*-\k** \divide\dn*2\relax
\divide\N*4\advance\N*-1\k**#3\def\xscale{#5}\def\yscale{#6}\relax
{\angle**0\rotate(#3)}\gmov*(#2,0){\sm*}\advance\k**\dn*
{\angle**0\rotate(\k**)}\gmov*(#2,0){\sl*}\advance\k**-\m**
\advance\l**\m**\loop\dt*-\dt*
\d*\d** \advance\d*\dt* \dd*\d** \advance\dd*1.3\dt*
\advance\k**\r*{\angle**0\rotate(\k**)}\rgmov*(\d*,0pt){\sl*}\relax
\advance\k**\r*{\angle**0\rotate(\k**)}\rgmov*(\dd*,0pt){\sl*}\relax
\advance\k**\r*{\angle**0\rotate(\k**)}\rgmov*(\d*,0pt){\sl*}\relax
\advance\k**\r*
\advance\N*-1\ifnum\N*>0\repeat\advance\k**\m**
{\angle**0\rotate(\k**)}\gmov*(#2,0){\sl*}\relax
{\angle**0\rotate(#4)}\gmov*(#2,0){\sl*}\def\xscale{1}\def\yscale{1}}}

\newcount\CatcodeOfAtSign
\CatcodeOfAtSign=\the\catcode`\@
\catcode`\@=11
\def\@arc#1[#2][#3]{\rlap{\Lengthunit=#1\Lengthunit
\sm*\l*arc(#2.1914,#3.0381)[#2][#3]\relax
\mov(#2.1914,#3.0381){\l*arc(#2.1622,#3.1084)[#2][#3]}\relax
\mov(#2.3536,#3.1465){\l*arc(#2.1084,#3.1622)[#2][#3]}\relax
\mov(#2.4619,#3.3086){\l*arc(#2.0381,#3.1914)[#2][#3]}}}

\def\dash@arc#1[#2][#3]{\rlap{\Lengthunit=#1\Lengthunit
\d*arc(#2.1914,#3.0381)[#2][#3]\relax
\mov(#2.1914,#3.0381){\d*arc(#2.1622,#3.1084)[#2][#3]}\relax
\mov(#2.3536,#3.1465){\d*arc(#2.1084,#3.1622)[#2][#3]}\relax
\mov(#2.4619,#3.3086){\d*arc(#2.0381,#3.1914)[#2][#3]}}}

\def\wave@arc#1[#2][#3]{\rlap{\Lengthunit=#1\Lengthunit
\w*lin(#2.1914,#3.0381)\relax
\mov(#2.1914,#3.0381){\w*lin(#2.1622,#3.1084)}\relax
\mov(#2.3536,#3.1465){\w*lin(#2.1084,#3.1622)}\relax
\mov(#2.4619,#3.3086){\w*lin(#2.0381,#3.1914)}}}

\def\bezier#1(#2,#3)(#4,#5)(#6,#7){\N*#1\l*\N* \advance\l*\*one
\d* #4\Lengthunit \advance\d* -#2\Lengthunit \multiply\d* \*two
\b* #6\Lengthunit \advance\b* -#2\Lengthunit
\advance\b*-\d* \divide\b*\N*
\d** #5\Lengthunit \advance\d** -#3\Lengthunit \multiply\d** \*two
\b** #7\Lengthunit \advance\b** -#3\Lengthunit
\advance\b** -\d** \divide\b**\N*
\mov(#2,#3){\sm*{\loop\ifnum\m*<\l*
\a*\m*\b* \advance\a*\d* \divide\a*\N* \multiply\a*\m*
\a**\m*\b** \advance\a**\d** \divide\a**\N* \multiply\a**\m*
\rmov*(\a*,\a**){\unhcopy\spl*}\advance\m*\*one\repeat}}}

\catcode`\*=12

\newcount\n@ast
\def\n@ast@#1{\n@ast0\relax\get@ast@#1\end}
\def\get@ast@#1{\ifx#1\end\let\next\relax\else
\ifx#1*\advance\n@ast1\fi\let\next\get@ast@\fi\next}

\newif\if@up \newif\if@dwn
\def\up@down@#1{\@upfalse\@dwnfalse
\if#1u\@uptrue\fi\if#1U\@uptrue\fi\if#1+\@uptrue\fi
\if#1d\@dwntrue\fi\if#1D\@dwntrue\fi\if#1-\@dwntrue\fi}

\def\halfcirc#1(#2)[#3]{{\Lengthunit=#2\Lengthunit\up@down@{#3}\relax
\if@up\mov(0,.5){\@arc[-][-]\@arc[+][-]}\fi
\if@dwn\mov(0,-.5){\@arc[-][+]\@arc[+][+]}\fi
\def\lft{\mov(0,.5){\@arc[-][-]}\mov(0,-.5){\@arc[-][+]}}\relax
\def\rght{\mov(0,.5){\@arc[+][-]}\mov(0,-.5){\@arc[+][+]}}\relax
\if#3l\lft\fi\if#3L\lft\fi\if#3r\rght\fi\if#3R\rght\fi
\n@ast@{#1}\relax
\ifnum\n@ast>0\if@up\shade[+]\fi\if@dwn\shade[-]\fi\fi
\ifnum\n@ast>1\if@up\dshade[+]\fi\if@dwn\dshade[-]\fi\fi}}

\def\halfdashcirc(#1)[#2]{{\Lengthunit=#1\Lengthunit\up@down@{#2}\relax
\if@up\mov(0,.5){\dash@arc[-][-]\dash@arc[+][-]}\fi
\if@dwn\mov(0,-.5){\dash@arc[-][+]\dash@arc[+][+]}\fi
\def\lft{\mov(0,.5){\dash@arc[-][-]}\mov(0,-.5){\dash@arc[-][+]}}\relax
\def\rght{\mov(0,.5){\dash@arc[+][-]}\mov(0,-.5){\dash@arc[+][+]}}\relax
\if#2l\lft\fi\if#2L\lft\fi\if#2r\rght\fi\if#2R\rght\fi}}

\def\halfwavecirc(#1)[#2]{{\Lengthunit=#1\Lengthunit\up@down@{#2}\relax
\if@up\mov(0,.5){\wave@arc[-][-]\wave@arc[+][-]}\fi
\if@dwn\mov(0,-.5){\wave@arc[-][+]\wave@arc[+][+]}\fi
\def\lft{\mov(0,.5){\wave@arc[-][-]}\mov(0,-.5){\wave@arc[-][+]}}\relax
\def\rght{\mov(0,.5){\wave@arc[+][-]}\mov(0,-.5){\wave@arc[+][+]}}\relax
\if#2l\lft\fi\if#2L\lft\fi\if#2r\rght\fi\if#2R\rght\fi}}

\catcode`\*=11

\def\Circle#1(#2){\halfcirc#1(#2)[u]\halfcirc#1(#2)[d]\n@ast@{#1}\relax
\ifnum\n@ast>0\L*=\xscale\Lengthunit
\ifnum\angle**=0\clap{\vrule width#2\L* height.1pt}\else
\L*=#2\L*\L*=.5\L*\special{em:linewidth .001pt}\relax
\rmov*(-\L*,0pt){\sm*}\rmov*(\L*,0pt){\sl*}\relax
\special{em:linewidth \the\linwid*}\fi\fi}

\catcode`\*=12

\def\wavecirc(#1){\halfwavecirc(#1)[u]\halfwavecirc(#1)[d]}

\def\dashcirc(#1){\halfdashcirc(#1)[u]\halfdashcirc(#1)[d]}

\def\xscale{1}
\def\yscale{1}

\def\Ellipse#1(#2)[#3,#4]{\def\xscale{#3}\def\yscale{#4}\relax
\Circle#1(#2)\def\xscale{1}\def\yscale{1}}

\def\dashEllipse(#1)[#2,#3]{\def\xscale{#2}\def\yscale{#3}\relax
\dashcirc(#1)\def\xscale{1}\def\yscale{1}}

\def\waveEllipse(#1)[#2,#3]{\def\xscale{#2}\def\yscale{#3}\relax
\wavecirc(#1)\def\xscale{1}\def\yscale{1}}

\def\halfEllipse#1(#2)[#3][#4,#5]{\def\xscale{#4}\def\yscale{#5}\relax
\halfcirc#1(#2)[#3]\def\xscale{1}\def\yscale{1}}

\def\halfdashEllipse(#1)[#2][#3,#4]{\def\xscale{#3}\def\yscale{#4}\relax
\halfdashcirc(#1)[#2]\def\xscale{1}\def\yscale{1}}

\def\halfwaveEllipse(#1)[#2][#3,#4]{\def\xscale{#3}\def\yscale{#4}\relax
\halfwavecirc(#1)[#2]\def\xscale{1}\def\yscale{1}}

\catcode`\@=\the\CatcodeOfAtSign
  
\begin{center}
{\bf PROTON POLARIZABILITY CORRECTION \\
TO THE HYDROGEN HYPERFINE SPLITTING}\footnote{Talk presented at the
Conference of RAS Nuclear Physics Department "Fundamental Interactions of
Elementary Particles" Moscow, ITEP, 16-20 November 1998}\\

\vspace{4mm}

R.N.~Faustov \\Scientific Council "Cybernetics" RAS\\
117333, Moscow, Vavilov, 40, Russia,\\
A.P.~Martynenko\\ Department of Theoretical Physics, Samara State University,\\
443011, Samara, Pavlov, 1, Russia,\\
V.A.~Saleev\\ Department of Theoretical Physics, Samara State University,\\
443011, Samara, Pavlov, 1, Russia
\end{center}  

\begin{abstract}
The contribution of $\Delta$ isobar to the correction on proton polarizability
in the hyperfine splitting of hydrogen and muonic hydrogen is calculated
with the account of the experimental data on $N-\Delta$ transition form factors.
\end{abstract}

\newpage

Investigation of the hyperfine splitting (HFS) of the hydrogen atom ground state
is considered during many years as a basic test of quantum electrodynamics \cite{BY}.
Experimental value of hydrogen hyperfine splitting was obtained with very
high accuracy \cite{HH}:
\begin{equation}
\Delta E^{hfs}_{exp}=1420405.7517667(9)~~KHz.
\end{equation}

Corresponding theoretical value of hydrogen hyperfine splitting may be written
at present time in the form:
\begin{equation}
\Delta E^{hfs}_{th}=\Delta E^F(1+\delta^{QED}+\delta^S+\delta^P),~~\Delta E^F=
\frac{8}{3}\alpha^4\frac{\mu_P m_p^2 m_e^2}{(m_p+m_e)^3},
\end{equation}
where $\mu_P$ is the proton magnetic moment,
$m_e$, $m_p$ are the masses of the electron and proton.
$\delta^{QED}$ denotes the contribution of higher-order quantum-
electrodynamical effects, which has the same form as for
muonium hyperfine structure. In the last years the theoretical accuracy of
this correction was essentially increased in \cite{KN1,KN2,EGS}.
Corrections $\delta^S$ and $\delta^P$ take into account the influence
of strong interaction. $\delta^S$ describes the effects of proton finite size
and recoil contribution.
$\delta^P$ is the correction of proton polarizability. Main uncertainty of
theoretical result (2) is connected with this term \cite{FG,VZ,BD,P,FP}.
The theoretical limitation for the proton polarizability contribution is
$|\delta^P| < 4 ppm$ \cite{SK}.
One of the main contributions to $\delta^P$ is determined by two photon diagrams,
when $\Delta$ - isobar or other baryon resonances may appear in the intermediate
states.
The amplitudes of virtual Compton scattering on electron and proton,
which appear in these diagrams, are not well known at present time.
Some estimation of $\delta^P$ was done in the paper \cite{ZSFCH}:
$\delta^P\sim 1\div 2 ppm$.
The correction to the hydrogen Lamb shift due to the proton electric and
magnetic polarizabilities was obtained in \cite{KS}.  
In this work we have suggested other approach to take into account resonance
intermediate states in $\delta^p$ correction of hydrogen HFS. Our approach is
based on using of transition form factors of nucleon to $\Delta$ isobar or
other baryon resonances. Such transition form factors may be measured
experimentally and then used for HFS calculation in hydrogen or muonic
hydrogen.

\begin{figure}
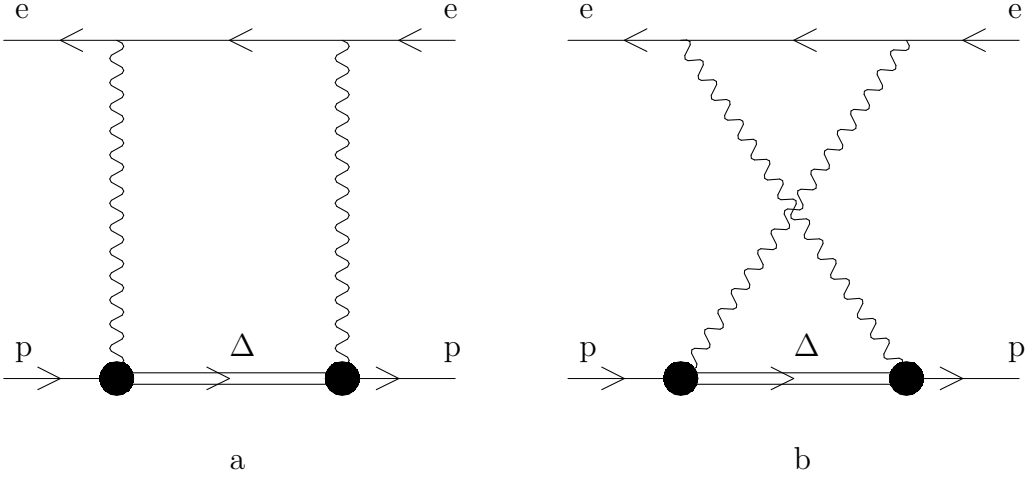

\magnitude=2000
\GRAPH(hsize=15){
\mov(0,0){\lin(1,0)}%
\mov(3,0){\lin(1,0)}%
\mov(0,3){\lin(4,0)}%
\mov(5,0){\lin(1,0)}%
\mov(8,0){\lin(1,0)}%
\mov(5,3){\lin(4,0)}%
\mov(3.,0.){\Circle**(0.3)}%
\mov(6.,0){\Circle**(0.3)}%
\mov(1.,0.){\Circle**(0.3)}%
\mov(8.,0.){\Circle**(0.3)}%
\mov(2.,-0.8){a}%
\mov(7.,-0.8){b}
\mov(0.5,3.){\lin(0.2,0.1)}%
\mov(0.5,3.){\lin(0.2,-0.1)}%
\mov(2.,3.){\lin(0.2,0.1)}%
\mov(2.,3.){\lin(0.2,-0.1)}%
\mov(3.5,3.){\lin(0.2,0.1)}%
\mov(3.5,3.){\lin(0.2,-0.1)}
\mov(5.5,3.){\lin(0.2,0.1)}%
\mov(5.5,3.){\lin(0.2,-0.1)}%
\mov(7.,3.){\lin(0.2,0.1)}%
\mov(7.,3.){\lin(0.2,-0.1)}%
\mov(8.5,3.){\lin(0.2,0.1)}%
\mov(8.5,3.){\lin(0.2,-0.1)}
\mov(0.5,0.){\lin(-0.2,0.1)}%
\mov(0.5,0.){\lin(-0.2,-0.1)}%
\mov(2.,0.){\lin(-0.2,0.1)}%
\mov(2.,0.){\lin(-0.2,-0.1)}%
\mov(3.5,0.){\lin(-0.2,0.1)}%
\mov(3.5,0.){\lin(-0.2,-0.1)}
\mov(5.5,0.){\lin(-0.2,0.1)}%
\mov(5.5,0.){\lin(-0.2,-0.1)}%
\mov(7.,0.){\lin(-0.2,0.1)}%
\mov(7.,0.){\lin(-0.2,-0.1)}%
\mov(8.5,0.){\lin(-0.2,0.1)}%
\mov(8.5,0.){\lin(-0.2,-0.1)}
\mov(1.,-0.05){\rectangle(2,0.1)}%
\mov(6.,-0.05){\rectangle(2,0.1)}%
\mov(1,0){\wavelin(0,3)}%
\mov(3,0){\wavelin(0,3)}
\mov(6,0){\wavelin(2,3)}
\mov(8,0){\wavelin(-2,3)}%
\mov(0.1,0.2){p}%
\mov(0.1,3.2){e}%
\mov(3.9,0.2){p}%
\mov(3.9,3.2){e}%
\mov(5.1,0.2){p}%
\mov(5.1,3.2){e}%
\mov(8.9,0.2){p}%
\mov(8.9,3.2){e}%
\mov(2.,0.2){$\Delta$}%
\mov(7.,0.2){$\Delta$}%
}
\caption{$\Delta$ isobar contribution to the hydrogen HFS}
\end{figure}  

Let consider the calculation of $\Delta$ isobar contribution to the hydrogen
HFS on the basis of quasipotential method \cite{MF}. The interaction operator
of electron and proton, corresponding to diagrams of figure, contains matrix
elements of electromagnetic current operator $<N^\ast(p_1)|J_\mu|N(p_2)>$
(N - nucleon, $N^\ast$ - nucleon resonance), depending on three relativistic
invariant form factors. Matrix element of the operator of electromagnetic
current for transitions  
$\frac{1}{2}^+\rightarrow\frac{3}{2}^-$, $\frac{5}{2}^+$, $\frac{7}{2}^-$, ...
takes the kind \cite{AR,JS}:
\begin{equation}
<N^\ast(p_2)|J_\mu|N(p_1)>=\bar\psi^{\mu_1...\mu_{j-1/2}}q_{\mu_2}...q_{\mu_{j-1/2}}
\end{equation}
\begin{displaymath}
[g_1(q^2)q_{\mu_1}(q_\mu q\cdot p_1-p_{1~\mu}q^2)
-2g_2(q^2)\varepsilon_{\mu_1\alpha\beta\gamma}p_{1~\alpha}q_\beta\varepsilon_
{\gamma\nu\lambda\mu}p_{1~\nu}q_\lambda+
\end{displaymath}
\begin{displaymath}
i(g_2(q^2)+g_3(q^2))Mq_{\mu_1}
\varepsilon_{\alpha\nu\lambda\mu}p_{1~\nu}q_\lambda\gamma_\alpha\gamma_5]u(p_1),
\end{displaymath}
where $\psi^{\mu_1...\mu_{j-1/2}}$ is the wave function of the baryon resonance,
$M$ is the mass of resonance. Relativistic invariant functions $g_i(q^2)$ (i=1, 2, 3)
may be called by electromagnetic nonelastic form factors.
To obtain the matrix element of electromagnetic current operator for
transitions $\frac{1}{2}^+\rightarrow \frac{3}{2}^+, \frac{5}{2}^-, $...
it is necessary to do the substitution $u(p_2)\rightarrow \gamma_5 u(p_2)$
in the expression (3).
Wave function of $\Delta$ isobar with $J^P=\frac{3}{2}^+$ is spin-vector
$\psi_\mu$. It satisfies to the relations:
$p_\mu\psi_\mu(p)=0$, $\gamma_\mu\psi_\mu(p)=0$, $(\hat p-M)=0$.
Using these equations we may write necessary matrix element of the
current $J_\mu$ in the form \cite{BR}:
\begin{equation}
<\Delta(p_2)|J_\mu|N(p_1)>=\bar\psi^\lambda T^{\lambda\mu}u(p-q),
\end{equation}
\begin{displaymath}
T^{\lambda\mu}=[G_1(q^2)(q^\lambda\gamma^\mu-
g^{\lambda\mu}\hat q)+
G_2(q^2)(q^\lambda P^\mu-g^{\lambda\mu}(qP))+  
G_3(q^2)(q^\lambda q^\mu-g^{\lambda\mu}q^2)]
\gamma_5
\end{displaymath}
where $p_2=p$, $p_1=p-q$, $P=p-\frac{1}{2}q$. Standard projector onto
the state with $J=\frac{3}{2}$ is
\begin{equation}
X^{\mu\nu}=\sum_\sigma\psi^\mu\bar\psi^\nu=(g^{\mu\nu}-\frac{1}{3}\gamma^\mu
\gamma^\nu+  
\frac{1}{3M}(p^\mu\gamma^\nu-p^\nu\gamma^\mu)-\frac{2}{3M^2}p^\mu
p^\nu)(\hat p+M).
\end{equation}

Functions $G_1(q^2)$, $G_2(q^2)$, $G_3(q^2)$ may be expressed in terms of more
usual set of form factors: electric $G_E$, magnetic $G_M$,
quadruple $G_C$:
\begin{equation}
G_1(s)=\frac{(G_M-G_E)M}{2[s+(M+m_p)^2]}
\end{equation}
\begin{equation}
G_2(s)=-\Biggl[G_M[(M-m_p)^2+s]+G_E(m_p^2+2m_pM-3M^2+s)-
\end{equation}
\begin{displaymath}
-2G_Cs\Biggr]\frac{1}{2[(M^2-m_p^2)^2+
s(s+2m_p^2+2M^2)]}
\end{displaymath}
\begin{equation}
G_3(s)=-\Biggl[G_M[(M-m_p)^2+s]+G_E((m_p+M)^2+4M^2+s)+
\end{equation}
\begin{displaymath}
+2G_C(m_p^2-M^2)\Biggr]\frac{1}{4[(M-
m_p)^2+s][(M+m_p)^2+s]},
\end{displaymath}
where $s=Q^2=-q^2$.

In the range of small $Q^2$ electromagnetic transition
$N\rightarrow\Delta (1232)$ may be considered as a magnetic dipole
transition $M_{1+}$. The contributions of electric quadruple
$E_{1+}$ and Coulomb quadruple $S_{1+}$ amplitudes are very small.
Recent experimental data in the range of $Q^2\approx 0$ are:
$E_{1+}/M_{1+}\sim$ -0.03, $S_{1+}/M_{1+}\sim$ -0.11 \cite{F,C,BE,S}.
As $Q^2$ increases the $N-\Delta$ transition form factors fall off more
like $Q^{-6}$ \cite{S}.  
The transition $\gamma^\ast p\rightarrow\Delta$
was studied on the basis of QCD sum rules in \cite{BR} and in the framework
of covariant diquark model in \cite{K}. Transition form factors $G_i(q^2)$
were calculated in \cite{BR} and it was shown good agreement of theoretical
results on form factors $G_M$, $G_E$, $G_C$ with experimental data.

Let consider the two photon amplitudes shown on figure. The electron factor
of one-loop amplitudes is equal to
\begin{equation}
M_e^{\mu\nu}=\bar v(p_1)\gamma^\mu\frac{(\hat p+\gamma_0 E_1+m_e)}{D_e(p)}
\gamma^\nu v(q_1),
\end{equation}
where $D_e(p)=p^2+2E_1p^0-\gamma^2$, $\gamma^2=E_1^2-m_e^2$.
The proton tensor of direct two photon diagram with $\Delta$ isobar in the
intermediate state may be written as follows:
\begin{equation}
M^{\mu\nu}_p=\bar u(q_2)T^{\mu\lambda}(p,q)\frac{X^{\lambda\omega}(p_2-p)}{D_\Delta(-p)}
T^{\omega\nu}(p,q)u(p_2).
\end{equation}
In the case of crossed diagram it is necessary to change
$\mu\Leftrightarrow \nu$ and $D_\Delta(-p)\rightarrow D_\Delta(p)$.
Choosing term $\sim \gamma^\mu\hat p\gamma^\nu$ in (9), we have projected
alternately electron and proton in the initial and final position on the
states with spin S=1 and S=0 by means of the operator  
\begin{equation}
\hat\Pi=u(p_2)\bar v(p_1)=\frac{1}{2\sqrt{2}}(1+\gamma^0)\hat\varepsilon
\end{equation}
($\varepsilon^\mu$ is the polarization vector of $^1S_3$ state, in the case of
$^1S_0$ it is necessary to change $\hat \varepsilon\rightarrow \gamma_5$),
in order to construct hyperfine part of the quasipotential. The expression
of two photon interaction operator, which appears after calculation of total
trace, contains along with different degrees of integration momentum p the
products of transition form factors $G_i(q^2)$. Further simplifications of
this quasipotential immediately follow from the above-mentioned experimental
data on the process $\gamma^\ast p\rightarrow \Delta$.
Considering N-$\Delta$ transition as a magnetic dipole process we may use
the following approximation:
\begin{equation}
G_1(s)=\frac{M}{2[s+(m_p+M)^2]}G_M(s),~G_2(s)=-\frac{G_1(s)}{M},  
G_3(s)=-\frac{G_1}{2M}.
\end{equation}
Expressing on this way all form factors through $G_1(q^2)$, we may represent
necessary quasipotentials of diagrams (a) and (b) in the kind:  
\begin{equation}
V^a_{2\gamma}=\frac{32(Z\alpha)^2}{9\pi^2m_p^2}\int \frac{d^4pG_1^2(p^2)}
{D_e(p)D_\Delta(-p)(p^2+i\epsilon)^2}  
[a_0+a_1p_0+a_2p_0^2+a_3p_0^3+a_4p_0^4]
\end{equation}
\begin{displaymath}
a_0=p^4[3p^2(2+\frac{m_p}{M})+4m_p^2-6m_pM-2m^2],
\end{displaymath}
\begin{displaymath}
a_1=3p^2[p^2(-\frac{m_p^2}{M}-12m_p-7M)-5m_p^3-5m_p^2M+2m_pM^2+2M^3],
\end{displaymath}
\begin{displaymath}
a_2=p^2[3p^2\left(\frac{11}{2}-\frac{m_p}{M}\right)+\frac{139}{2}m_p^2+
63m_pM-\frac{11}{2}M^2],
\end{displaymath}
\begin{displaymath}
a_3=3[p^2(\frac{m_p^2}{M}-15m_p)-7m_p^3-7m_p^2M],~~~a_4=21m_p^2,
\end{displaymath}
\begin{equation}
V^b_{2\gamma}=\frac{32(Z\alpha)^2}{9\pi^2m_p^2}\int \frac{d^4pG_1^2(p^2)}
{D_e(p)D_\Delta(p)(p^2+i\epsilon)^2}  
[b_0+b_1p_0+b_2p_0^2+b_3p_0^3+b_4p_0^4]
\end{equation}
\begin{displaymath}
b_0=p^4[3p^2(2+\frac{m_p}{M})+4m_p^2-6m_pM-2m^2],
\end{displaymath}
\begin{displaymath}
b_1=3p^2[p^2(\frac{m_p^2}{M}+2m_p-5M)-m_p^35m_p^2M+4m_pM^2+4M^3],
\end{displaymath}
\begin{displaymath}
b_2=p^2[3p^2\left(-\frac{7}{2}-\frac{m_p}{M}\right)-\frac{59}{2}m_p^2+
27m_pM+\frac{7}{2}M^2],
\end{displaymath}
\begin{displaymath}
b_3=3[p^2(-\frac{m_p^2}{M}-11m_p)-5m_p^3-5m_p^2M],~~~b_4=-15m_p^2,
\end{displaymath}
where we have set also the relative motion particle momenta $\vec p=\vec q=0$.
Let transform (13)-(14) to the kind, which is convenient for numerical
integration \cite{MF1}. Factor $\frac{1}{D_\Delta(-p)D_e(p)}$ of direct two photon
diagram may be written as follows:  
\begin{equation}
\frac{1}{D_\Delta(-p)D_e(p)}=\frac{1}{2(m_p+m_e)\left[p^2-\frac{m_e
(M^2-m_p^2)}{m_p}+i\varepsilon\right]}  
\left[\frac{2m_p}{D_\Delta(-p)}+\frac{2m_e}{D_e(p)}\right].
\end{equation}
Other $\Delta$ isobar term $\frac{1}{D_\Delta}(p)$, which is in (14) may be
represented in the form
\begin{equation}
\frac{1}{D_\Delta(p)D_e(p)}=\frac{1}{2(m_p-m_e)\left(p_0-
\frac{M^2-m_p^2}{2m_p}+
i\varepsilon\right)}  
\left[\frac{1}{D_e(p)}-\frac{1}{D_\Delta(p)}\right].
\end{equation}
After that we rotate the $p_0$ contour to the imaginary axis and call the
new variable $p_0=i\xi$. The integration factors are modified as follows:
\begin{equation}
\frac{1}{p^2+i\varepsilon}\rightarrow -\frac{1}{\vec p^2+\xi^2},  
\frac{1}{D_\Delta(p)}\rightarrow -\frac{\vec p^2+\xi^2+M^2-m_p^2+
2im_p\xi}{(\vec p^2+\xi^2+M^2-m_p^2)^2+4m_p^2\xi^2}.
\end{equation}
For the calculation of (13)-(14) we have used the expression (12) for form
factor $G_1$ and the relation of magnetic form factor, obtained in \cite{BR}:
\begin{equation}
G_M(\vec p^2,\xi^2)=\frac{2s_0^3S_0^3}{l_Nl_\Delta(Q^2+s_0+S_0)^3}  
\frac{1}{(1-3\sigma+(1-\sigma)\sqrt{1-4\sigma})},
\end{equation}
where $\sigma=s_0S_0/(Q^2+s_0+S_0)^2$, $l^2_N$=$s_0^3/12$, $l^2_\Delta$=
$S_0^3/10$. Numerical values of parameters $s_0$, $S_0$ were taken also as in
\cite{BR}: $s_0=2.3~GeV^2$, $S_0=3.5~GeV^2$. After averaging $V_{2\gamma}^a$ and
$V_{2\gamma}^b$ over Coulomb wave functions we can write the contribution of
$\Delta$ isobar to the proton polarizability correction $\delta^P$ of hydrogen
atom as a sum of two integrals:  
\begin{equation}
\Delta E_1^{hfs}=-\frac{E_F\alpha m_em_p}{3\pi^2}\int_0^\infty
\frac{p^2 dpd\xi}{(\vec p^2+\xi^2)^2[\vec p^2+\xi^2+(M+m_p)^2]^2}\times
\end{equation}
\begin{displaymath}
\times\frac{G_M^2(\vec p^2,\xi^2)}{[(\vec p^2+\xi^2+M^2-m_p^2)^2+
4m_p^2\xi^2][(\vec p^2+\xi^2)^2+4m_e^2\xi^2]}\times
\end{displaymath}
\begin{displaymath}
[(\vec p^2+\xi^2)(\vec p^2+\xi^2+M^2-m_p^2)+4m_em_p\xi^2](a_0-a_2\xi^2+
a_4\xi^4)+
\end{displaymath}
\begin{displaymath}
+2\xi^2[m_p(\vec p^2+\xi^2)-m_e(\vec p^2+\xi^2+M^2-m_p^2)](a_1-a_3\xi^2)],
\end{displaymath}

\begin{equation}
\Delta E_2^{hfs}=-\frac{E_F\alpha m_em_p}{3\pi^2}\int_0^\infty
\frac{p^2 dpd\xi}{(\vec p^2+\xi^2)^2[\vec p^2+\xi^2+(M+m_p)^2]^2}
\end{equation}
\begin{displaymath}
\frac{G_M^2(\vec p^2,\xi^2)}{[(\vec p^2+\xi^2+M^2-m_p^2)^2+
4m_p^2\xi^2][(\vec p^2+\xi^2)^2+4m_e^2\xi^2]}\times
\end{displaymath}
\begin{displaymath}
[(\vec p^2+\xi^2)(\vec p^2+\xi^2+M^2-m_p^2)-4m_em_p\xi^2](b_0-b_2\xi^2+
b_4\xi^4)+
\end{displaymath}
\begin{displaymath}
-2\xi^2[m_p(\vec p^2+\xi^2)+m_e(\vec p^2+\xi^2+M^2-m_p^2)](b_1-b_3\xi^2)].
\end{displaymath}

Numerical value of considered contribution to $\delta^p$ is equal to
\begin{equation}
\Delta E^{hfs}_p (\Delta-isobar)=-0.12\cdot 10^{-6} \Delta E^F.
\end{equation}
The result of our calculation is in good agreement with the estimation
of resonance J= $\frac{3}{2}$ contribution to HFS, obtained in \cite{FG,VZ}.
In the case of muonic hydrogen the relative value of this contribution to
HFS run to (-27 ppm). Theoretical error, not exceeding $10\%$, is connected
with the used approximation in equations (19)-(20) (the contribution of
form factors $G_E$ and $G_C$ was omitted). It must be emphasized, that all
transition form factors fall off like $Q^{-6}$ with a rise in $Q^2$ and the
value of integrals (19)-(20) is determined by the range of small $Q^2\sim m_e^2\alpha^2$.
So the form factor $G_M(Q^2)$ behaviour in the vicinity of such quantities
$Q^2$ is mainly responsible for the value of correction $\Delta E_p^{hfs}$. A
simple rise of resonance mass in (19)-(20) even with the using of form factor (18)
leads to drastic decreasing of correction (21) in modulus: at M=1.5 GeV
$\Delta E_p^{hfs}$ =-0.04 ppm,
and at M=1.7 Gev $\Delta E_p^{hfs}$ =-0.004 ppm. So for calculation of resonance
contribution to the proton polarizability correction we may restrict only by
some low-lying nucleon resonances. In this approach definite consideration
of other baryonic resonances in the intermediate state is possible after
theoretical study of corresponding form factors or with the availability of
necessary experimental data.

We are grateful to Karshenboim S.G., Khriplovich I.B., Petrunkin V.A., Sen'kov R.A. for useful
discussions. This work was supported by Russian
Foundation for Basic Research (grant no 98-02-16185), and the Program
"Universities of Russia - Fundamental Researches" (grant no 2759).

\end{document}